# Potentiodynamic Electrochemical Impedance Spectroscopy of Silver on Platinum in Underpotential and Overpotential Deposition

Genady A. Ragoisha[*], Alexander S. Bondarenko

*Physico-Chemical Research Institute, Belarusian State University, Minsk 220050, Belarus*

**Abstract**

Simultaneous monitoring of ac and dc responses of the electrode-electrolyte interface with potentiodynamic electrochemical impedance spectroscopy (PDEIS) in silver underpotential and overpotential deposition on platinum has confirmed the role of intrinsic Pt surface changes in the irreversibility of Ag underpotential deposition and disclosed exceptionally high stability of Ag monolayer on Pt. PDEIS has been demonstrated to be a convenient means for wet surface chemistry monitoring.

**Keywords**

Electrochemical methods, Electrical transport measurements, Crystallization, Surface chemical reaction, Surface electrical transport, Silver, Metal-electrolyte interfaces, Adatoms, Virtual instruments

**Introduction**

Potentiodynamic Electrochemical Impedance Spectroscopy (PDEIS) acquires ac and dc responses of an electrochemical interface in a single potential scan and analyses the ac response in terms of equivalent electric circuit parameters dependences on the potential [1,2]. This is especially helpful in investigations of short-lived states of variable interfaces that cannot endure sequential application of several techniques for comprehensive in situ electrochemical characterisation of the interface transient states in various transformations. PDEIS has provided in situ investigation of anion adsorption during copper monolayer formation on gold [2] and electrochemical characterisation of metal monolayers during their formation and destruction on metallic and nonmetallic supports [1-3]. The investigation of Cu underpotential deposition (upd) on Au in presence of anions [2] has disclosed irreversibility of constituent processes in the reaction that appeared to be reversible from voltammograms. In the present work we have applied PDEIS to investigate a transition from upd to overpotential deposition (opd) in silver cathodic deposition on Pt. Irreversibility of Ag upd on Pt manifested itself already in cyclic voltammetry [4,5], which was ascribed to oxidative-reductive [5] and structural [6] changes of Pt surface layer, so this system required an in situ technique for characterisation of transient states. As we will show later, the theory predicts the potentiodynamic ac response of the irreversible upd to be significantly different from that of the reversible upd. We have confirmed the theoretical considerations by the experimental investigation of silver upd using PDEIS and demonstrated high sensitivity of PDEIS to variation in the upd-opd system status.

[*] Corresponding author
E-mail: ragoishag@bsu.by
Fax: 375-17-2264696

**Theory**

A linear component of the ac electrochemical response of a upd at a perturbation of the electrode potential $\Delta E = \bar{E} \exp(j\omega t)$ may be presented in the following functional form [7,8]:

$$\Delta i = \left(\frac{\partial i}{\partial E}\right) \Delta E + \left(\frac{\partial i}{\partial C_s}\right) \Delta C_s + \left(\frac{\partial i}{\partial \theta}\right) \Delta \theta \qquad (1)$$

where $\Delta i$, $\Delta C_s$, $\Delta \theta$ are the respective variations of the current, metal cation surface concentration and surface coverage of metal adatoms formed as a result of upd, $\bar{E}$ is the perturbation amplitude, $j$ – imaginary unit, $\omega$ – angular frequency, t – time. The dependence of the current on these variables comes from the kinetic equations:

$$i = nF[k_f C_s(1-\theta) - k_b\theta] \quad (2)$$

$$k_f = k_0\exp[-\alpha nF\eta/RT] \quad (3)$$

$$k_b = k_0\exp[(1-\alpha)nF\eta/RT] \quad (4)$$

where $k_0$ is a standard rate constant, $k_f$ and $k_b$ – constants of respective forward and backward reactions, R – gas constant, T – absolute temperature, F – Faraday constant, η – overpotential, n – number of electrons involved in the electrochemical reaction, α - transfer coefficient.

Though actual responses may contain also nonlinear components in Eq. (1), nonlinear contribution has proved to be usually insignificant at a low perturbation amplitude practised in electrochemical impedance spectroscopy. Nonlinear effects that might arise from inter-frequency interactions in multi-frequency probing were excluded in this work by using a sequential probing with streams of synchronised wavelets generated by virtual instruments [1].

The first term in Eq.(1) manifests itself in the active impedance component represented in equivalent circuits by a charge transfer resistance $R_{ct}$, the second one depends on diffusion, thus it can show up either in Warburg impedance or other equivalent circuit fragments that take account of local concentration changes at different frequencies of perturbation. The last term should be significantly different in reversible and irreversible upd. In the case of reversible upd, a separate quasi-equilibrium can be established at each potential in the upd range with characteristic coverage of adatoms [9]. The perturbation of the potential results in the perturbation of θ manifested in pseudocapacitance [8,9]. There may be even several pseudocapacitances in the equivalent circuit, if a monolayer formation is accompanied by adsorption effects. The pseudocapacitances variations with the potential were used to study the dynamics of Cu monolayer formation and conjugated anion adsorption with PDEIS [2]. On the contrary, in irreversible upd there is no characteristic coverage and, consequently, no oscillation of θ to produce the monolayer pseudocapacitance via the last term of Eq. (1). Ag monolayer formation and destruction on Pt proceed in different potential ranges, so their ac responses are also separated on the potential scale and can be disclosed separately by a charge transfer pseudoresistance and the diffusion impedance attributed to $Ag^+$. Also analytical descriptions of the charge transfer pseudoresistance should be different in these two cases, because of the lack of characteristic quasi-equilibrium coverage at each potential and a dependence of $\theta$ on time of cathodic deposition. The $R_{ct}$ of the reversible upd tends to measure the exchange current density at the corresponding characteristic coverage with the approach to the stationary condition [9], while the EEC parameters of the irreversible upd should be less attached to the potential and show greater history dependence.

**Experimental**

PDEIS spectra were recorded and analysed in terms of equivalent circuits with a virtual PDEIS spectrometer – a computer program that used a common potentiostat as the actuator in the electrochemical response probing. The description of PDEIS is posted on Chemweb [1], so we omit it to save the area for some comments on the peculiarities of PDEIS relative to common electrochemical impedance spectroscopy (EIS).

Unlike common EIS that works with stationary systems in wide frequency ranges, PDEIS works with non-stationary systems and has to limit the frequency range, so that the potential variation in the scan is low compared to one resulting from the ac probing at a lowest frequency. For that reason, a low frequency range is available only at slow scans. Though the frequency range restriction may result in a loss of equivalent circuit elements that respond outside the probed frequency range, the ability of PDEIS to decompose an ac response into the responses of different elements of the equivalent electric circuit (EEC) in the range embraced by the probing is much higher than of a common EIS, as the potential variation restricts considerably the choice of EEC. To achieve high accuracy of fitting we recorded the EIS spectra with a low increment on the potential scale (1 or 2 mV), so that the scan steps were lower than the probing amplitude and the neighbouring 2D slices of the PDEIS spectra could be compared for noise detection and subtraction. In this work we have used an updated spectrometer program (version 1.4) that implemented a digital subtraction of 50 Hz noise just in the measurement stage by a superposition of the responses recorded at different phase shifts of the wavelets in the probing signal stream. This was important as the informative frequency range of Ag/Pt electrochemical ac response embraced this frequency. Practically, 18 frequencies (from 877 Hz to 32 Hz) were found to be sufficient for smooth processing of the ac response both in the upd and opd ranges.

Polycrystalline platinum wire (purchased from ChemPur) was flame annealed in each experiment and after cooling in air placed into 1 mM $AgNO_3$ + 0.1M $HNO_3$ electrolyte deareated with nitrogen. The geometrical area of the Pt electrode was 0.015 $cm^2$. A platinized Pt counter electrode of high surface area was used to eliminate the counter electrode impedance from the total impedance. The potential was controlled versus a Ag|AgCl|KCl(sat) reference electrode. In order to prevent the contamination of the working solution by chloride, the reference electrode was placed in a tap-isolated compartment of the three-electrode electrochemical cell.

**Results and discussion**

Fig. 1 shows a PDEIS spectrum for upd and opd of Ag on Pt at cyclic potential scan reversed from the cathodic to anodic direction at 200 mV, with cyclic voltammograms corresponding to different reversal potentials. The fitting procedure for several constant potential sections of the spectra is illustrated in Fig. 2 and the dependencies of the EEC parameters on the potential at different reversal potentials and scan rates, obtained from different PDEIS spectra, are shown in Figs. 3-5.

The EEC obtained by the PDEIS spectrometer fitting routine appeared to be relatively simple (Fig. 2a). As predicted from the theoretical considerations, it does not contain capacitance of adsorption. A constant phase element (CPE) represents the double layer capacitance with a phase shift slightly below $\pi/2$ radians (n factor varied from 0.92 to 1.0). The imperfection of the double layer capacitance is common for electrochemical systems and in the present case the nonstationarity of the system was the additional reason for such behaviour. We are leaving this point for a consideration in a special publication. A simple Warburg element described well the diffusion impedance. This was somewhat unexpected, as the $Ag^+$ diffusion to metal nuclei could be more complex than the one considered in the semi-infinite diffusion model of the Warburg element. Probably, the complexity of the diffusion impedance was masked by the truncation of a low frequency range.

Both the ac and dc responses show the irreversible Ag upd ($C_M$ –cathodic formation of a monolayer, $A_M$ – anodic destruction of a monolayer) and the opd ($C_B$ – cathodic bulk Ag deposition, $A_B$ – anodic oxidation of bulk Ag). Pt electrode pseudocapacitance Q shows a hysteresis in a cyclic scan in absence of $Ag^+$ (dotted line in Fig. 3a) that corresponds to irreversible changes in the surface layer. Ag upd manifests itself by a Q rise in the cathodic scan between 500 and 350 mV and the subsequent Q decrease on the reverse scan above 700 mV. Ag upd range overlaps with the range of intrinsic changes in Pt surface layer. That is obviously the reason of the high irreversibility of silver redox transformations in the monolayer.

A shift of the reversal potential in the cyclic scan between the upd to the opd ranges (Fig. 3a) results in a Q decrease in the beginning of bulk Ag growth. This could be due either to a double layer expansion or dielectric permittivity decrease in the nucleation stage. On the reverse scan Q shows a small maximum at the potential of bulk Ag growth initiation, fading with the reversal potential shift into the opd range, and a high peak slightly shifted to anodic potentials from the anodic current maximum. The positive potential slope of this peak is less steep than of the current peak on the voltammogram (Fig. 3b). This shows that the status of the double electric layer typical for Ag monolayer recovers upon bulk Ag oxidation with a delay, however, the similarity of Q(E) in the upd range at different reversal potential in Fig. 3a makes evidence of a monolayer preservation during the bulk silver cathodic deposition and anodic oxidation. The inverses of $R_{ct}$ and Warburg constant A in Fig 4 correlate with the dc current, thus confirming the belonging of the corresponding EEC elements to the Faraday impedance. We shall not consider in detail these two parameters in this paper.

With a scan rate increase, the Q increase, attributed to the monolayer formation, embraces a smaller potential range, and the anodic peak, attributed to bulk silver oxidation, becomes less steep, shifts to higher potentials and finally interflows with the upd plateau that rises with the scan rate (Fig. 5). So the history of the system shows itself in the strain of the potential dependences of the ac response.

Thus, the simultaneous monitoring of ac and dc responses in Ag upd and opd on Pt with PDEIS gives the opportunity to control the surface status with a number of variable parameters of equivalent electric circuits related with different processes on the interface.

**Conclusions**

The investigation of Ag upd and opd on Pt with PDEIS has confirmed the role of intrinsic changes of Pt surface layers in the irreversibility of Ag monolayer deposition. The preservation of the double layer pseudocapacitace dependence on the potential in the upd range after bulk silver cathodic deposition and anodic oxidation has uncovered exceptionally high stability of Ag monolayer. The multi-parameter monitoring of nonstationary interfaces provided by PDEIS appears to be a convenient means of dynamic surfaces control and this may be of interest for nanotechnological and nanochemical processes that employ wet surfaces.

**References**

[1] G.A. Ragoisha and A.S. Bondarenko, Solid State Phenom. 90-91, 103 (2003); electronic version: http://preprint.chemweb.com/physchem/0301002.

[2] G.A. Ragoisha and A.S. Bondarenko, Electrochem. Commun. 5 (2003) 392.

[3] G.A. Ragoisha and A.S. Bondarenko, *Physics, Chemistry and Application of Nanostructures*, World Scientific, 2003, 373; electronic version: http://preprint.chemweb.com/physchem/0301005.

[4] E. Herrero, L. J. Buller, H. D. Abruna, Chem. Rev., 101 (2001) 1897.

[5] A. Vaskevich, E. Gileadi, J. Electroanal. Chem. 442 (1998) 147.

[6] E. D. Mishina, N. Ohtaa, Q.K. Yua, S. Nakabayashia, Surf. Sci. Lett. 494 (2001) 748.

[7] A. Lasia, Electrochemical Impedance Spectroscopy and its Applications, http://alfa.chem.uw.edu.pl/LasiaLecture/EIS1.pdf.

[8] A. Lasia, Applications of Electrochemical Impedance Spectroscopy to Hydrogen adsorption, Evolution and Absorption into Metals, http://alfa.chem.uw.edu.pl/LasiaLecture/EIS2.pdf.

[9] S. Morin, H. Dumont, B.E. Conway, J. Electroanal. Chem. 412 (1996) 39.

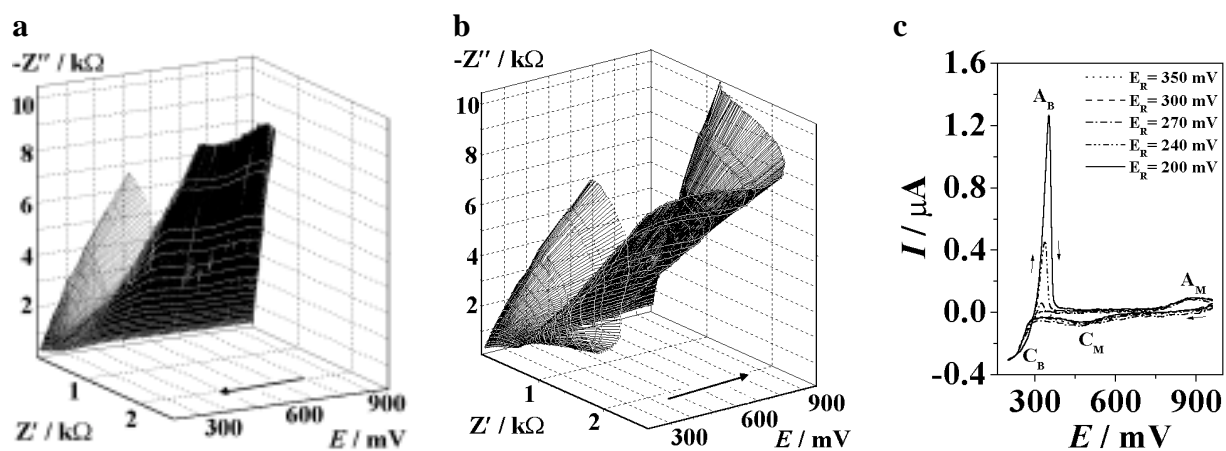

Fig. 1. Ag upd on Pt. (a) Cathodic and (b) anodic branch of cyclic PDEIS spectrum; (c) cyclic voltammograms with different reversal potentials ($E_R$). $dE/dt$ = 2.3 mV/s. Electrolyte: 1mM $AgNO_3$ + 0.1M $HNO_3$. Scan direction is shown by arrows.

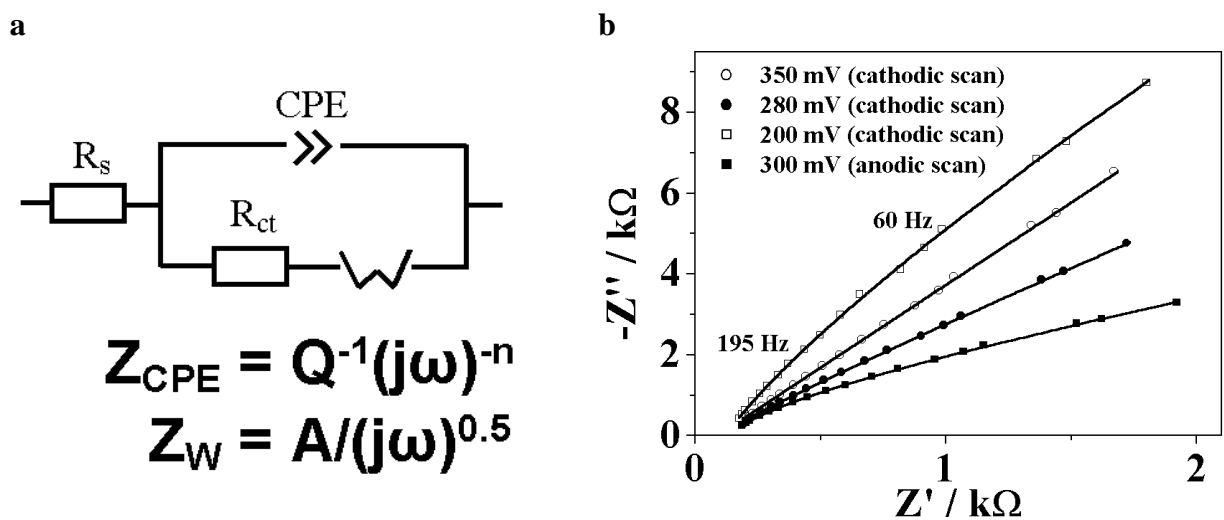

Fig. 2. (a) Equivalent electric circuit for the PDEIS spectrum shown in Fig.1; (b) Slices of the PDEIS spectrum shown in Fig. 1 (Nyquist plots) at different electrode potential with experimental points and simulated curves.

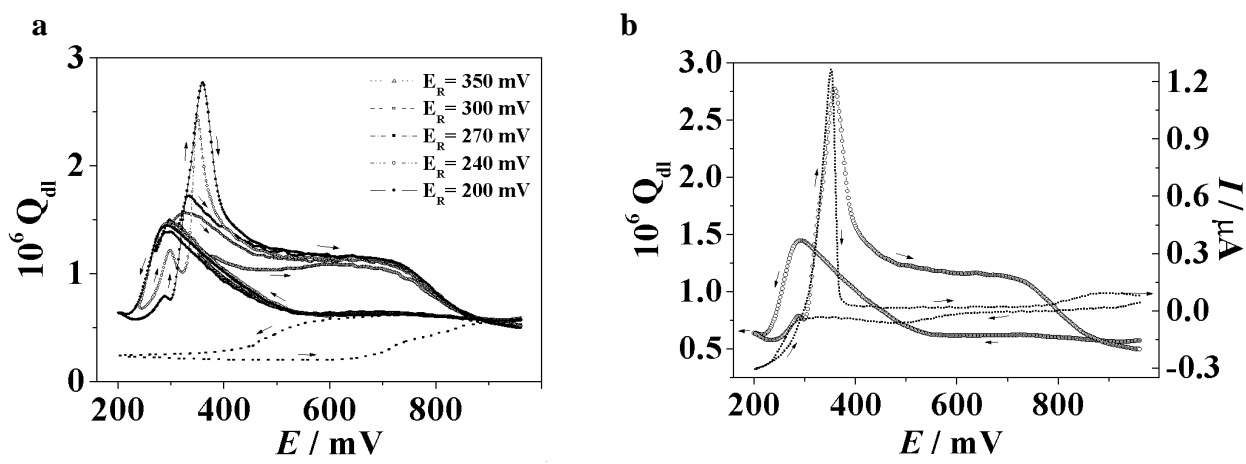

Fig. 3. (a) Double layer pseudocapacitance dependencies on electrode potential at a cyclic scan with different reversal potential $E_R$. Dotted line characterise the substrate in absence of $Ag^+$. (b) Double layer pseudo-capacitance dependence on potential (circles) and the corresponding cyclic voltammogram (dotted line) at $E_R$=200mV. $dE/dt$ = 2.3 mV/s.

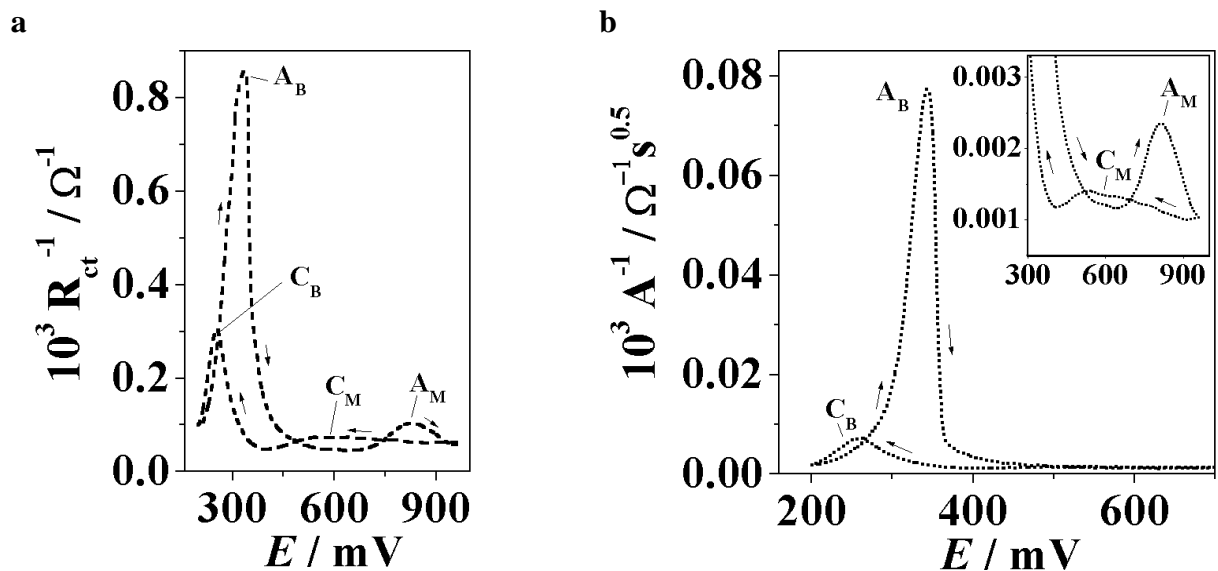

Fig. 4. Inverses of (a) charge transfer resistance and (b) Warburg coefficient as functions of the potential in a cyclic scan with $E_R$=200mV. The insert shows the upd peaks at extended scale.

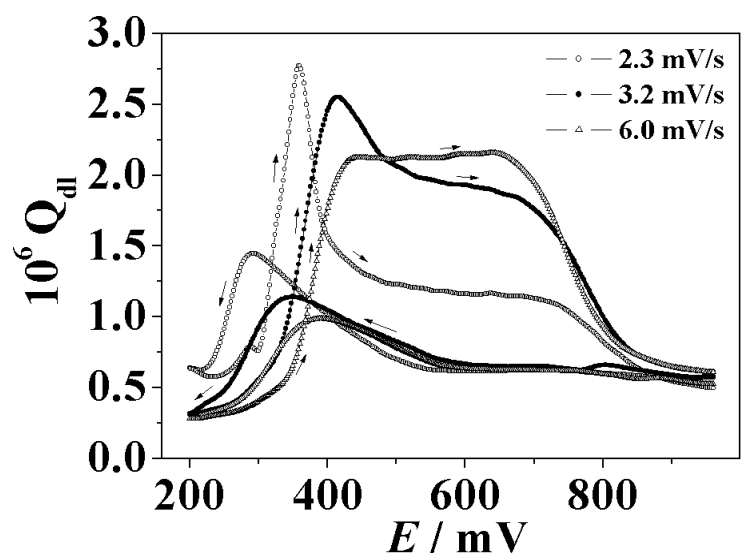

Fig. 5. Double layer pseudocapacitance dependencies for Pt electrode in 1mM $AgNO_3$ + 0.1M $HNO_3$ at different rates of cyclic potential scan.